# Opportunistic Scheduling in Underlay Cognitive Radio based MIMO-RF/FSO Networks

Neeraj Varshney, *Student Member, IEEE*, Prabhat K. Sharma, *Senior Member, IEEE*, and Mohamed-Slim Alouini, *Fellow, IEEE*

*Abstract*—This work proposes an optimal metric for opportunistic scheduling of secondary user transmitters (SU-TXs) in underlay cognitive radio based multiple-input multiple-output radio frequency/free space optical (MIMO-RF/FSO) decode-and-forward system with fixed and proportional interference power constraints. To analyze the performance of the proposed system, the closed-form expressions are derived for the exact and asymptotic outage probabilities considering orthogonal space-time block coded transmission over Nakagami-$m$ fading RF links. Further, the FSO link between the relay and destination eNodeB is modeled as the Generalized Málaga ($\mathcal{M}$) turbulence channel with pointing errors. Finally, simulation results are presented to develop several interesting insights into the end-to-end system performance and the selection probabilities of SU-TXs. It is also shown that the proposed system outperforms the ones existing in the current literature.

## I. INTRODUCTION

In the recent advances of wireless and optical communication technologies, the mixed radio frequency and free-space optical (RF/FSO) systems have gained significant popularity towards improving the communication reliability, data rates, network connectivity etc. Due to above advantages along with lower costs and deployment times, the mixed RF/FSO systems have been explored in various applications including cognitive radios (CRs) [1]–[5]. In CRs, specifically in the underlay mode, the mixed RF/FSO systems not only provide the traditional RF/FSO advantages [6] but also avoid the RF interference from the relay to primary user receiver (PU-RX) that exists in RF/RF systems.

### A. Related work

The authors in [1]–[3] considered the amplify-forward (AF) relaying at the intermediate node with single-input single-output (SISO) RF links, whereas [4] employed decode-and-forward (DF) relaying with multiple-input multiple-output (MIMO) RF links while analyzing the performance of mixed RF/FSO underlay CR networks. In contrast to [1]–[4], the work in [5] recently addressed the problem of the secondary user transmitter (SU-TX) selection and analyzed the outage performance of a multi-user mixed underlay RF/FSO network with multiple destination nodes. However, the selection metric proposed therein is suboptimal and results in considerably poor performance at the destination node. Further, the work in [5] does not exploit the transmit diversity that can significantly enhance the system performance with limited transmit power arises due to interference constraint in underlay CR mode. Further, it is also worth noting that the AF protocol considered at the relay in [1]–[3], [5] is not easily realizable in practice since both the RF and FSO links operate at different frequencies. The relay therefore has to decode the information symbol prior to modulation over an optical carrier, as shown in [4].

### B. Contributions

To the best of our knowledge, none of the works in the existing literature proposed the optimal metric for SU-TX selection in a MIMO-RF/FSO underlay CR network employing both the fixed and proportional power constraints. With this motivation, this work analyzes the end-to-end outage performance of MIMO-RF/FSO underlay CR network considering the optimal metric for opportunistic scheduling. In contrast to AF protocol [5], the novel closed-form expressions are derived for the exact and asymptotic outage probabilities at the cognitive base station (eNodeB) with orthogonal space-time block coded (OSTBC) based transmission and DF relaying at the SU-TX and relay, respectively. The RF links are assumed to be Nakagami-$m$ fading in nature, whereas the FSO link is modeled as the generalized Málaga ($\mathcal{M}$)-distributed atmospheric turbulence channel with zero-boresight misalignment errors. It is worth noting that Gamma-Gamma distributed FSO link in [4] is a special case of $\mathcal{M}$-distribution considered in this work. Simulations results are finally presented to develop several interesting insights into the system performance. Moreover, it is also demonstrated that the proposed system outperforms the ones existing in the current literature.

### C. Organization

The organization of the rest of the paper is as follows. The underlay CR based MIMO-RF/FSO system model and the scheduling scheme considering multiple SU-TXs are presented in Section II. The analytical framework for analyzing the exact and asymptotic outage behavior is provided in Section III. Section IV presents the numerical results obtained analytically and through simulations. Finally the paper is concluded in Section V.

Neeraj Varshney is with the Department of Electrical Engineering and Computer Science, Syracuse University, Syracuse, New York 13244, USA (e-mail:nvarshne@syr.edu).

Prabhat K. Sharma is with the Department of Electronics and Communication Engineering, Visvesvaraya National Institute of Technology (NIT) Nagpur, Maharastra 440010, India (e-mail:prabhatsharma@ece.vnit.ac.in).

Mohamed-Slim Alouini is with the Computer, Electrical and Mathematical Science and Engineering Division, King Abdullah University of Science and Technology (KAUST), Thuwal 23955-6900, Saudi Arabia (e-mail: slim.alouini@kaust.edu.sa).

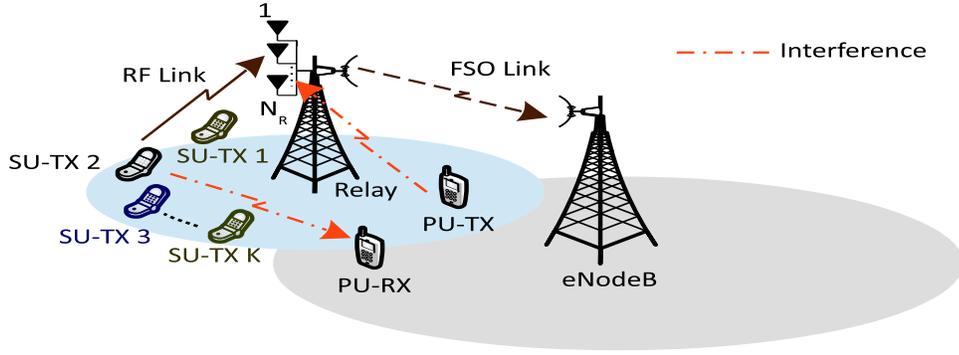

Fig. 1. Schematic diagram of a underlay CR based mixed RF/FSO DF relay network with multiple SU-TXs.

## II. COGNITIVE SYSTEM MODEL WITH SU-TX SELECTION

Consider an underlay based cognitive scenario with opportunistic scheduling as shown in Fig. 1, where one out of $K$ SU-TXs is opportunistically selected to transmit the information to the cognitive eNodeB through RF/FSO link in the presence of a primary transmitter and receiver pair. It is worth noting that since only one SU-TX transmits at a given time using the spectral resource of the PU-RX, the relay $R$ does not experience any interference except from the primary user transmitter (PU-TX). Moreover, to limit the interference at the PU-RX, the selected SU-TX adaptively controls its transmit power using the fixed and proportional interference power constraints. Further, this work considers that each SU-TX and PU-RX are equipped with $N_S$ and $N_P$ antennas, respectively. The intermediate node $R$ possesses $N_R$ antennas to receive the OSTBC transmission from the selected SU-TX over MIMO RF link, and subsequently forwards the decoded symbols to the destination eNodeB in sequential manner through an FSO link after modulating over an optical carrier. It is important to note that due to the orthogonal nature of the effective channel matrix over the different transmit time instants, a simple decorrelator can serve as the optimal ML detector at $R$.

Corresponding to the transmission of an OSTBC block $\mathbf{X}_{SR}^{(k)} \in \mathbb{C}^{N_S \times T}$ by the $k$th SU-TX using the transmit power $P_S^k$, the received codeword $\mathbf{Y}_{SR}^{(k)} \in \mathbb{C}^{N_R \times T}$ at $R$, is given as

$$\mathbf{Y}_{SR}^{(k)} = \sqrt{\frac{P_S^k}{R_c N_S}} \mathbf{H}_{SR}^{(k)} \mathbf{X}_{SR}^{(k)} + \mathbf{W}_{SR}^{(k)}, \quad (1)$$

where $T$ represents the block length and $R_c = \frac{B}{T}$ denotes the rate of the OSTBC where $B$ symbols are encoded in a block. The entries of SU TX-$R$ MIMO channel matrix $\mathbf{H}_{SR}^{(k)} \in \mathbb{C}^{N_R \times N_S}$ are statistically independent and identically distributed (i.i.d.) Nakagami-$m$ random channel coefficients with variance $\delta_{SR,k}^2$ and severity parameter $m_{SR,k}$ each. Similar to [7], the entries of the additive white noise plus primary user interference matrix $\mathbf{W}_{SR}^{(k)} \in \mathbb{C}^{N_R \times T}$ are i.i.d. symmetric complex Gaussian with power $\eta_0$. For OSTBC transmission, the instantaneous signal-to-noise ratio (SNR) per symbol at $R$ can be obtained as [8]

$$\gamma_{SR}^{(k)}[i] = \frac{P_S^k ||\mathbf{H}_{SR}^{(k)}||_F^2}{R_c N_S \eta_0}, \quad (2)$$

where $i = 1, 2, \cdots, B$. In underlay cognitive scenario, the SU-TX transmits with power

$$P_S^k = \begin{cases} P_M^k & \text{if } ||\mathbf{H}_{SP}^{(k)}||_F^2 \leq \frac{P_A}{P_M^k}, \\ \frac{P_A}{||\mathbf{H}_{SP}^{(k)}||_F^2} & \text{otherwise,} \end{cases}$$

$$= \min\left\{ P_M^k, \frac{P_A}{||\mathbf{H}_{SP}^{(k)}||_F^2} \right\},$$

to ensure that the interference at the PU-RX is within the specified threshold $P_A$. Here $P_M^k$ is the maximum available power at the $k$th SU-TX, $\mathbf{H}_{SP}^{(k)} \in \mathbb{C}^{N_P \times N_S}$ denotes the MIMO channel matrix between the $k$th SU-TX and PU-RX, that comprises of i.i.d. Nakagami-$m$ distributed random variables with variance $\delta_{SP,k}^2$ and severity parameter $m_{SP,k}$. Substituting the aforementioned $P_S^k$ in (2), the instantaneous SNR $\gamma_{SR}^{(k)}[i]$ at $R$ can be written as

$$\gamma_{SR}^{(k)}[i] = \frac{1}{R_c N_S \eta_0} \min\left\{ P_M^k ||\mathbf{H}_{SR}^{(k)}||_F^2, \frac{P_A ||\mathbf{H}_{SR}^{(k)}||_F^2}{||\mathbf{H}_{SP}^{(k)}||_F^2} \right\}. \quad (3)$$

Using the above expression, the optimal metric $\beta^*$ for choosing the SU-TX for transmission is given as

$$\beta^* = \max_{k=1,2,\cdots,K} \min\left\{ P_M^k G_{SR}^{(k)}, \frac{P_A G_{SR}^{(k)}}{G_{SP}^{(k)}} \right\}, \quad (4)$$

where $G_{SR}^{(k)} = ||\mathbf{H}_{SR}^{(k)}||_F^2$ and $G_{SP}^{(k)} = ||\mathbf{H}_{SP}^{(k)}||_F^2$ denote the gain of the cognitive SU TX-relay and cross SU TX-PU RX links respectively. The instantaneous SNR $\gamma_{SR}^*[i]$ at $R$ with opportunistic scheduling can now be expressed as

$$\gamma_{SR}^*[i] = \frac{\beta^*}{R_c N_S \eta_0}. \quad (5)$$

## III. OUTAGE PERFORMANCE ANALYSIS

To analyze the end-to-end outage performance, this section first derives the cumulative distribution functions (CDFs) of the RF and FSO link SNRs.

### A. RF Link Statistics

The CDF of the instantaneous SNR $\gamma_{SR}^*[i]$ given in (5) can be obtained as

$$F_{\gamma_{SR}^*[i]}(x) = \Pr(\beta^* \leq R_c N_S \eta_0 x) = F_{\beta^*}(R_c N_S \eta_0 x), \quad (6)$$

where $F_{\beta^*}(\cdot)$ denotes the CDF of the selection metric $\beta^*$ and can be derived using (4) as

$$F_{\beta^*}(x) = \Pr\left(\max_{k=1,2,\cdots,K} \min\left\{P_M^k G_{SR}^{(k)}, \frac{P_A G_{SR}^{(k)}}{G_{SP}^{(k)}}\right\} \leq x\right)$$

$$= \prod_{k=1}^{K} \Pr\left(\min\left\{P_M^k G_{SR}^{(k)}, \frac{P_A G_{SR}^{(k)}}{G_{SP}^{(k)}}\right\} \leq x\right)$$

$$= \prod_{k=1}^{K} F_\zeta(x), \quad (7)$$

where $\zeta \triangleq \min\left\{P_M^k G_{SR}^{(k)}, \frac{P_A G_{SR}^{(k)}}{G_{SP}^{(k)}}\right\}$ and $F_\zeta(x)$ denotes the CDF of $\zeta$, and is derived in Appendix A as

$$F_\zeta(x) = \frac{1}{\Gamma(\tau_2)}\left[\frac{1}{\Gamma(\tau_1)}\gamma\left(\tau_1, \frac{x m_{SR,k}}{P_M^k \delta_{SR,k}^2}\right)\gamma\left(\tau_2, \frac{P_A m_{SP,k}}{P_M^k \delta_{SP,k}^2}\right)\right.$$
$$+ \Gamma\left(\tau_2, \frac{P_A m_{SP,k}}{P_M^k \delta_{SP,k}^2}\right) - \sum_{l=0}^{\tau_1-1}\frac{1}{l!}\left(\frac{x m_{SR,k}}{P_A \delta_{SR,k}^2}\right)^l$$
$$\times \left(\frac{\delta_{SP,k}^2}{m_{SP,k}}\right)^l \times \left(1 + \frac{x m_{SR,k}\delta_{SP,k}^2}{P_A m_{SP,k}\delta_{SR,k}^2}\right)^{-\tau_2-l}$$
$$\left.\times \Gamma\left(\tau_2 + l, \left(\frac{m_{SP,k}}{\delta_{SP,k}^2} + \frac{x m_{SR,k}}{P_A \delta_{SR,k}^2}\right)\frac{P_A}{P_M^k}\right)\right], \quad (8)$$

where $\tau_1 = m_{SR,k} N_S N_R$ and $\tau_2 = m_{SP,k} N_S N_P$. The quantities $\gamma(\cdot,\cdot)$ and $\Gamma(\cdot,\cdot)$ denote the lower and upper incomplete Gamma functions respectively [9, Eqs. (8.350.1), (8.350.2)].

### B. FSO Link Statistics

The FSO channel $h_{RD}[i]$ between $R$ and eNodeB incorporates the effects of atmospheric turbulence induced fading and misalignment fading, and is given by $h_{RD}[i] = h_a[i]h_m[i]$, where $h_a[i]$ and $h_m[i]$ are the coefficients for turbulence induced and misalignment fading, respectively. The fading coefficient $h_a[i]$ follows the Málaga ($\mathcal{M}$)-distribution that physically includes three components: first, an line of sight (LOS) component $U_L$; second, a component $U_S^C$ which is coupled to the LOS component $U_L$ and scattered by the eddies on the propagation axis, and the third component $U_S^G$ which is scattered by the eddies off the propagation axis. The probability density function (PDF) of $h_a[i]$ is given as [10]

$$f_{h_a[i]}(h_a) = \chi \sum_{k=1}^{\beta} a(k) h_a^{\frac{\alpha+k}{2}-1} K_v\left(2\sqrt{\frac{\alpha\beta h_a}{\tau\beta+\Omega'}}\right), \quad (9)$$

where $K_v(\cdot)$ is the modified Bessel function of second kind with order $v = \alpha - k$ [9]. The positive parameter $\alpha$ represents the effective number of large scale cells of the scattering process, and $\beta$ is a natural number that represents amount of turbulence induced fading[1]. Further, the various quantities in above equation are defined as, $\chi = \frac{2\alpha^{\frac{\alpha}{2}}}{\tau^{1+\frac{\alpha}{2}}\Gamma(\alpha)}\left(\frac{\tau\beta}{\tau\beta+\Omega'}\right)^{\beta+\frac{\alpha}{2}}$, $\tau = 2b_0(1-\rho)$,

[1]The parameter $\beta$ can also be considered as a real number but that results in an infinite summation in the PDF.

$a(k) = \binom{\beta-1}{k-1}\frac{(\tau\beta+\Omega')^{1-\frac{k}{2}}}{(k-1)!}\left(\frac{\Omega'}{\tau}\right)^{k-1}\left(\frac{\alpha}{\beta}\right)^{\frac{k}{2}}$, $\Omega' = \Omega + 2\rho b_0 + 2\sqrt{2b_0\rho\Omega}\cos(\varphi_1-\varphi_2)$. Here the term $2b_0$ is the average power of total scattered component, i.e., $(U_S^G + U_S^C)$ and the parameter $0 \leq \rho \leq 1$ represents the amount of scattering power coupled to the LOS component $U_S^C$. The term $\Omega$ is the average power of the LOS component $U_L$, $\varphi_1$ and $\varphi_2$ are the deterministic phases of LOS component $U_L$ and coupled to LOS component $U_S^C$, respectively.

For misalignment errors, the PDF of fading coefficient $h_m[i]$ is given by [11]

$$f_{h_m[i]}(h_m) = \frac{\zeta^2}{(A_0)^{\zeta^2}}(h_m)^{\zeta^2-1}, \quad 0 \leq h_m \leq A_0, \quad (10)$$

where $\zeta = \frac{w_{zeq}}{2\sigma_s}$, is the ratio between the equivalent beam width and jitter standard deviation at the eNodeB, which is a measure of severity of misalignment error effect. The parameter $A_0 = [\text{erf}(v)]^2$ is a constant quantity that defines the pointing loss, where $\text{erf}(\cdot)$ is the error function and $v = \frac{\sqrt{\pi}a}{\sqrt{2}w_z}$, where $a$ and $w_z$ denote the radius of the detection aperture and the beam waist respectively.

The PDF of channel coefficient $h_{RD}[i] = h_a[i]h_m[i]$ can be calculated as

$$f_{h_{RD}[i]}(h) = \int_{h/A_0}^{\infty} f_{h_a[i]}(h_a)\, f_{h_{RD}[i]|h_a[i]}(h|h_a)\, dh_a. \quad (11)$$

Using (9), (10) and (11), the PDF of channel coefficient $h_{RD}[i]$ can be obtained as

$$f_{h_{RD}[i]}(h)$$
$$= \frac{\zeta^2\chi}{2h}\sum_{k=1}^{\beta} b_k\, \mathrm{G}_{1,3}^{3,0}\left(\frac{\alpha\beta h}{A_0(\tau\beta+\Omega')}\middle|\begin{array}{c}\zeta^2+1\\ \zeta^2, \alpha, k\end{array}\right), \quad (12)$$

where $b_k = a_k\left[\frac{\alpha\beta}{(\tau\beta+\Omega')}\right]^{-(\alpha+k)/2}$ and $\mathrm{G}_{p,q}^{m,n}\left(x\middle|\begin{array}{c}a_1,\cdots,a_p\\ b_1,\cdots,b_q\end{array}\right)$ is the Meijer's G function [9, Eq. 9.301].

The instantaneous SNR at the destination eNodeB corresponding to transmission over the FSO link can be given as $\gamma_{RD}[i] = \frac{(\eta_e h_{RD}[i])^r}{N_0}$, where $r$ is the detection technique parameter, $r = 1$ for heterodyne detection, $r = 2$ for IM/DD detection, $\eta_e$ and $N_0$ denote the effective photoelectric conversion ratio and AWGN sample at eNodeB respectively.

With some mathematical manipulations, one can easily derive the CDF of $\gamma_{RD}[i]$ using (12) as [12]

$$F_{\gamma_{RD}[i]}(x) = \frac{\zeta^2\chi}{(2r)(2\pi)^{r-1}}\sum_{k=1}^{\beta} b_k r^{\alpha+k-1}$$
$$\times \mathrm{G}_{r+1,3r+1}^{3r,1}\left(\frac{B^r}{r^{2r}\bar{\gamma}_{RD}}x\middle|\begin{array}{c}1,\kappa_1\\ \kappa_2, 0\end{array}\right), \quad (13)$$

where the average SNR is given as $\bar{\gamma}_{RD} = \frac{(A_0\eta_e\zeta^2(\tau+\Omega'))^r}{N_0(1+\zeta^2)^r}$ and the various terms are defined as, $B = \frac{\zeta^2\alpha\beta(\tau+\Omega')}{(1+\zeta^2)(\tau\beta+\Omega')}$, $\kappa_1 = \frac{\zeta^2+1}{r},\cdots,\frac{\zeta^2+r}{r}$ consists of $r$ terms, and $\kappa_2 = \frac{\zeta^2}{r},\cdots,\frac{\zeta^2+r-1}{r},\frac{\alpha}{r},\cdots,\frac{\alpha+r-1}{r},\frac{m}{r},\cdots,\frac{m+r-1}{r}$ consists of $3r$ terms.

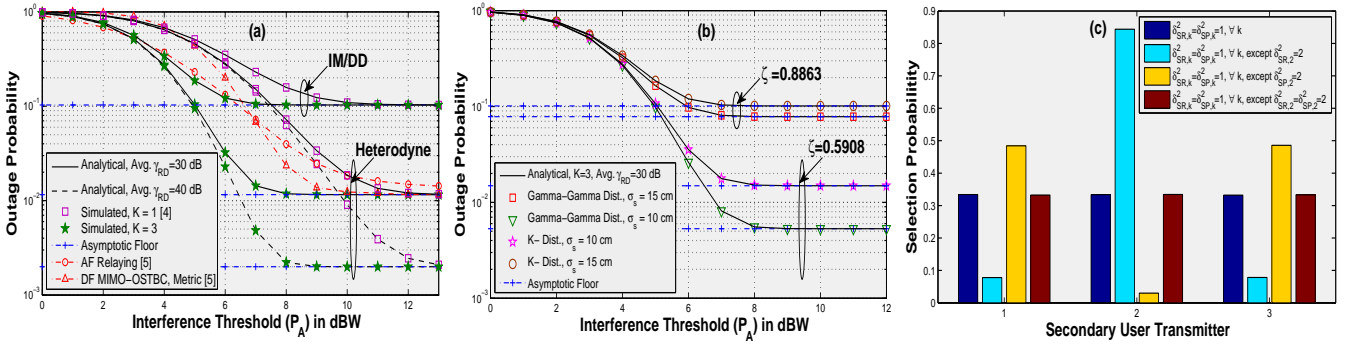

Fig. 2. With SU-TX selection; (a) Outage probability versus interference threshold $P_A$ with fixed $\bar{\gamma}_{RD} \in \{30, 40\}$ dB and $\mathcal{M}(\alpha = 2.296, \beta = 2, \gamma = 0.0872, \rho = 0.596, \Omega' = 1.085)$ fading FSO link with pointing error $\zeta = 0.8863$ (b) Outage probability versus interference threshold $P_A$ with $\bar{\gamma}_{RD} = 30$ dB and $\mathcal{M}(\alpha = 8, \beta = 4, \gamma = 0, \rho = 1, \Omega' = 1)$ for Gamma-Gamma fading FSO link and $\mathcal{M}(\alpha = 8, \beta = 4, \gamma = 0.2158, \rho = 0, \Omega' = 0)$ for K-distributed fading FSO link with pointing error $\zeta \in \{0.5908, 0.8863\}$ (c) Selection Probabilities of SU-TXs under various channel conditions with $P_A = 20$ dBW and Nakagami parameters $m_{SR,k} = m_{SP,k} = 2.5 \forall k$.

## C. Exact and Asymptotic Outage Probabilities

For a given SNR outage threshold $\gamma_{th}$, the end-to-end outage probability in underlay CR based MIMO-OSTBC RF/FSO DF system with opportunistic scheduling can be expressed as

$$P_{\text{out}}(\gamma_{\text{th}}) = \frac{1}{B} \sum_{i=1}^{B} \Pr(\min\{\gamma^*_{SR}[i], \gamma_{RD}[i]\} \leq \gamma_{\text{th}}), \quad (14)$$

where

$$\Pr(\min\{\gamma^*_{SR}[i], \gamma_{RD}[i]\} \leq \gamma_{\text{th}}) = 1 - (1 - F_{\gamma^*_{SR}[i]}(\gamma_{\text{th}}))(1 - F_{\gamma_{RD}[i]}(\gamma_{\text{th}})), \quad (15)$$

where the expressions for $F_{\gamma^*_{SR}[i]}(\gamma_{th})$ and $F_{\gamma_{RD}[i]}(\gamma_{th})$ are given in (6) and (13) respectively. Further, using the above expression and considering the fact that the Meijer's G function in (13) reduces to zero as $\bar{\gamma}_{RD}$ approaches $\infty$, the asymptotic floor when $\bar{\gamma}_{RD} \to \infty$ can be obtained as

$$\lim_{\bar{\gamma}_{RD} \to \infty} P_{\text{out}}(\gamma_{\text{th}}) = F_{\gamma^*_{SR}[i]}(\gamma_{\text{th}}). \quad (16)$$

On the other hand, the asymptotic floor for the scenarios when $P_A \to \infty$ and a fixed value of $\bar{\gamma}_{RD}$ is given as

$$\lim_{P_A \to \infty} P_{\text{out}}(\gamma_{\text{th}}) = 1 - (1 - F_{\gamma^*_{SR}[i]}(\gamma_{\text{th}}))(1 - F_{\gamma_{RD}[i]}(\gamma_{\text{th}})), \quad (17)$$

where $F_{\gamma^*_{SR}[i]}(\gamma_{th})$ can be derived for $P_A \to \infty$ as

$$F_{\gamma^*_{SR}[i]}(\gamma_{\text{th}}) = \prod_{k=1}^{K} \frac{1}{\Gamma(\tau_1)} \gamma\left(\tau_1, \frac{R_c N_S \eta_0 \gamma_{\text{th}} m_{SR,k}}{P_M^k \delta_{SR,k}^2}\right). \quad (18)$$

## IV. SIMULATION RESULTS AND INFERENCE

For simulations, we consider the presence of $K = 3$ SU-TXs, each employs rate $\frac{1}{2} \mathcal{G}_C^3$ OSTBC for transmission using $N_S = 3$ antennas. The other parameters are considered as, $N_R = N_P = 2$ antennas at R and PU-RX, SNR outage threshold $\gamma_{th} = 3$ dB, noise power $\eta_0 = 1$, maximum transmit power $P_M^k = 27$ dBW $\forall k$, and average channel gains $\delta_{SR,k}^2 = \delta_{SP,k}^2 = 1, \forall k$ with Nakagami parameters $m_{SR,k} = m_{SP,k} = 1.5, \forall k$ unless otherwise stated. For FSO link, link length $L = 1$ km, wavelength $\lambda = 785$ nm, and all other parameters e.g., refraction structure parameter $C_n^2$ are considered similar to the ones given in [12].

Fig. 2(a) shows the outage probability versus the interference threshold $P_A$ considering both IM/DD and heterodyne detection at eNodeB. Firstly, it can be observed that the opportunistic scheduling of $K = 3$ SU-TXs using (4) significantly improves the system performance in comparison to the scenario when only one SU-TX is present [4]. The improvement in low and moderate values of $P_A$ arises due to the increase in diversity order of the RF transmission by selecting one best out of 3 SU-TXs. However, the system experiences the identical asymptotic floor (17) for both $K = 3$ with opportunistic selection and $K = 1$ [4] scenarios. This is owing to the fact that the end-to-end performance of the system at high $P_A$ dominates by the weak FSO link. Further, the outage floor decreases as the average SNR $\bar{\gamma}_{RD}$ of FSO link increases. Moreover, one can also observe that the heterodyne detection at eNodeB outperforms the IM/DD based detection.

In contrast to [4], the analytical outage results are applicable for various fading scenarios such as Gamma-Gamma, K-distribution etc. [10, Table 1] in FSO link, as can be seen in Fig. 2(b). Further, one can also notice therein that the outage probability increases as the pointing error displacement standard deviation (jitter) at eNodeB increases. Fig. 2(a) also compares the performance of proposed SU-TX selection metric with the existing work in [5]. For fair comparison, we employed the selection metric proposed in [5] in MIMO-OSTBC RF/FSO DF system. It can be clearly seen in Fig. 2(a) that the proposed selection metric outperforms the one considered in [5]. This is owing to the fact that the metric in [5] is sub-optimal since the selection is done by only maximizing the channel gains between SU-TXs and relay, and after selection, the power is regulated by dividing $P_A$ with maximum of the channel gains of SU-TXs and PU-RX. One can also note that the proposed MIMO-OSTBC RF/FSO system with DF relaying protocol achieves low outage probability values in comparison to [5] that considered single antenna based SU-TXs and AF relaying at the relay. However, there is a marginal gap in the performance at low values of $P_A$, which arises due to power normalization factor $R_C N_S$ in MIMO-OSTBC based transmission.

This work also develops several interesting insights into the selection probabilities of SU-TXs, as can be seen in Fig. 2(c). Firstly, one can observe that for the scenarios when each SU-TX is located at equal distances from $R$ and PU-Rx, i.e., $\delta_{SP,k}^2 = \delta_{SR,k}^2 = 1, \forall k$, each SU-TX has equal probability of being selected for transmission. For the scenario, when 2nd SU-TX is closer to the $R$ than the PU-RX, i.e., $\delta_{SP,k}^2 = \delta_{SR,k}^2 = 1, \forall k$ except $\delta_{SR,2}^2 = 2$, the system enhances the end-to-end performance by choosing 2nd SU-TX approximately 84% of the times. On the other hand, when 2nd SU-TX is close to the PU-RX in comparison to the $R$, i.e., $\delta_{SP,k}^2 = \delta_{SR,k}^2 = 1, \forall k$ except $\delta_{SP,2}^2 = 2$, the probability of choosing 2nd SU-TX for transmission reduces to approximately 3% that in turns increases the selection probabilities of other SU-TXs. However, if the 2nd SU-TX is located close to the $R$ in comparison with the other SU-TXs and also have equal distances from $R$ and PU-RX, i.e., $\delta_{SP,k}^2 = \delta_{SR,k}^2 = 1, \forall k$ except $\delta_{SR,2}^2 = \delta_{SP,k}^2 = 2$, each of the SU-TXs will have approximately equal chance for transmission.

## V. CONCLUSION

In this paper, an optimal metric for opportunistically scheduling of multiple SU-TXs is proposed in an underlay CR network considering both the fixed and proportional interference power constraints. To analyze the performance of proposed scheduling scheme in underlay CR based MIMO-RF/FSO system, exact and asymptotic closed-form expressions are derived for the outage probability considering the OSTBC based transmission between each SU-TX and relay nodes. It has been shown that the proposed system achieves low outage probability values in comparison to the several existing works.

## APPENDIX A
### DERIVATION FOR $F_\zeta(x)$

The CDF $F_\zeta(x)$ of $\zeta = \min\left\{P_M^k G_{SR}^{(k)}, \frac{P_A G_{SR}^{(k)}}{G_{SP}^{(k)}}\right\}$ can be derived as

$$F_\zeta(x) = \int_0^{\frac{P_A}{P_M^k}} \left(\int_0^{\frac{x}{P_M^k}} f_{G_{SR}^{(k)}}(y) dy\right) f_{G_{SP}^{(k)}}(z) dz$$
$$+ \int_{\frac{P_A}{P_M^k}}^{\infty} \left(\int_0^{\frac{xz}{P_A}} f_{G_{SR}^{(k)}}(y) dy\right) f_{G_{SP}^{(k)}}(z) dz$$
$$= F_{G_{SR}^{(k)}}\left(\frac{x}{P_M^k}\right) F_{G_{SP}^{(k)}}\left(\frac{P_A}{P_M^k}\right)$$
$$+ \underbrace{\int_{\frac{P_A}{P_M^k}}^{\infty} F_{G_{SR}^{(k)}}\left(\frac{xz}{P_A}\right) f_{G_{SP}^{(k)}}(z) dz}_{\triangleq I}, \quad (19)$$

where $F_{G_{SR}^{(k)}}(\cdot)$ denotes the CDF of $G_{SR}^{(k)} = ||\mathbf{H}_{SR}^{(k)}||_F^2$ and is given as

$$F_{G_{SR}^{(k)}}(x) = \frac{1}{\Gamma(\tau_1)} \gamma\left(\tau_1, \frac{x m_{SR,k}}{\delta_{SR,k}^2}\right), \quad (20)$$

where $\tau_1 = m_{SR,k} N_S N_R$ and $\gamma(\cdot, \cdot)$ is the lower incomplete Gamma function [9, Eq. (8.350.1)]. The quantities $F_{G_{SP}^{(k)}}(\cdot)$ and $f_{G_{SP}^{(k)}}(\cdot)$ in (19) denote the CDF and PDF of $G_{SP}^{(k)} = ||\mathbf{H}_{SP}^{(k)}||_F^2$ respectively and are given as

$$F_{G_{SP}^{(k)}}(x) = \frac{1}{\Gamma(\tau_2)} \gamma\left(\tau_2, \frac{x m_{SP,k}}{\delta_{SP,k}^2}\right), \quad (21)$$

$$f_{G_{SP}^{(k)}}(x) = \frac{(m_{SP,k}/\delta_{SP,k})^{\tau_2}}{\Gamma(\tau_2)} x^{\tau_2 - 1} \exp\left(-\frac{x m_{SP,k}}{\delta_{SP,k}^2}\right), \quad (22)$$

where $\tau_2 = m_{SP,k} N_S N_P$. Employing the expressions (20), (22) and subsequently using the identities $\gamma(s,x) = \Gamma(s) - \Gamma(s) \exp(-x) \sum_{l=0}^{s-1} \frac{x^l}{l!}$ and $\int_x^\infty t^{s-1} \exp(-\mu t) dt = \frac{\Gamma(s, \mu x)}{\mu^s}$ from [9, Eqs. 8.352.1, 3.351.2], the integral $I$ in (19) can be solved as

$$I = \frac{1}{\Gamma(\tau_2)} \left[ \Gamma\left(\tau_2, \frac{P_A m_{SP,k}}{P_M^k \delta_{SP,k}^2}\right) - \sum_{l=0}^{\tau_1 - 1} \frac{1}{l!} \left(\frac{x m_{SR,k}}{P_A \delta_{SR,k}^2}\right)^l \right.$$
$$\times \left(\frac{\delta_{SP,k}^2}{m_{SP,k}}\right)^l \left(1 + \frac{x m_{SR,k} \delta_{SP,k}^2}{P_A m_{SP,k} \delta_{SR,k}^2}\right)^{-\tau_2 - l}$$
$$\left. \times \Gamma\left(\tau_2 + l, \left(\frac{m_{SP,k}}{\delta_{SP,k}^2} + \frac{x m_{SR,k}}{P_A \delta_{SR,k}^2}\right) \frac{P_A}{P_M^k}\right) \right], \quad (23)$$

where $\Gamma(\cdot, \cdot)$ is the upper incomplete Gamma function [9, Eq. (8.350.2)]. Finally, substituting the above expression along with (20), (21) for $F_{G_{SR}^{(k)}}(x)$ and $F_{G_{SP}^{(k)}}(x)$ in (19), the closed-form expression for the CDF $F_\zeta(x)$ can be derived as (8).